\documentclass[aps,prd,twocolumn,groupedaddress,floatfix]{revtex4}

\usepackage{graphicx}

\begin{document}


\title{Characterizing a cosmic string with the statistics of string lensing}

\author{Masamune Oguri$^{1}$ and Keitaro Takahashi$^{2}$}
\affiliation{
$^1$Department of Astrophysical Sciences, Princeton University, Princeton, NJ 08544. \\
$^2$Department of Physics, Princeton University, Princeton,  NJ 08544.
}

\date{\today}

\begin{abstract}
The deep imaging of the field of an observed lensing event by a cosmic
string reveals many additional lensing events. We study the statistics
of such string lensing. We derive explicit expressions for the
distributions of image separations of lensing by a cosmic string and
point out that they are quite sensitive to parameters which characterize
the cosmic string, such as the redshift and tension of the cosmic string.  
Thus the statistics of string lensing events add new important
information on the cosmic string which cannot be obtained from the
detailed investigation of one lensing event.  
\end{abstract}


\maketitle

\section{Introduction}

Cosmic strings are linear topological defects of a gauge field
which may have been formed in symmetry-breaking phase transition
in the early universe. They are characterized by the tension $\mu$,
which is of order the square of the symmetry-breaking scale.
Their gravitational effect is measured by a dimensionless parameter
$G\mu/c^{2}$. It is about $10^{-6}$ for GUT (grand unified theory) strings.
Cosmic strings had been studied as an alternative to inflation as a
means of generating the primordial fluctuation. However, recent
precise observations of the cosmic microwave background by Wilkinson
Microwave Anisotropy Probe and large scale structure by Sloan Digital
Sky Survey have excluded the possibility of their
dominant contribution to the primordial fluctuations. These
observations give constraints on the dimensionless parameter which are
typically $G\mu/c^{2} \lesssim 10^{-6} - 10^{-7}$
\cite{pogosian03,pogosian04,jeong05,lo05}. For recent reviews, see
\cite{kibble04,polchinski04,perivolaropoulos05}.

Despite the severe constraints, there has been a remarkable revival
of interest. On the theoretical side, the possibility of superstrings
of cosmological scale and their observability has been discussed by
several authors especially in the context of brane inflation
\cite{jones03,dvali04,copeland04}. They have almost the same observational
consequences as the conventional cosmic strings. A striking feature of
cosmic superstrings is their reconnection probability which differs
from that of the conventional cosmic strings by many orders. Thus the
difference may allow us to distinguish them observationally
\cite{jackson04}. Also it was argued in \cite{jeannerot03} that cosmic
strings should be present in a wide class of grand unified
inflationary models.

On the observational side, a possible detection of a cosmic string was
recently made by Sazhin et al. \cite{sazhin03}. They claimed that a pair
of galaxies CSL-1 is a galaxy at $z=0.46$ gravitationally lensed by a
foreground cosmic string. The observed image separation $1.9''$
implies that somewhat large tension is needed, but this does not pose
any serious problem if e.g., one considers cosmic strings with high
winding numbers \cite{donaire05}. The string lens hypothesis is still
to be explored, 
but if it is the case the system serves as an ideal laboratory to
study very early universe. As emphasized by \cite{huterer03}, a key
observation to distinguish string lensing from the chance projection
effect is deep wide-field imaging to identify other nearby lensing
events (see also \cite{shlaer05} for cosmic string lensing). In this
paper, we study how such lensing events constrain the properties of
the cosmic string. Specifically, we compute analytically the
distribution of image separations and show that it is sensitive to
parameters which characterize the string, such as the redshift, the
tension of the cosmic string and its coherence length. Throughout the
paper, we assume a Lambda-dominated cosmology with $\Omega_M=0.3$,
$\Omega_\Lambda=0.7$, and $h=H_0/100{\rm km\,s^{-1}Mpc^{-1}}=0.7$, and
also assume that cosmic strings are non-relativistic.

\section{Formulation}
We consider a situation that background galaxies at redshift $z_s$ are
multiply imaged by a cosmic string at redshift $z_l$. The tension of
the string is denoted by $\mu$. The angular separation between multiply
imaged galaxies is given as \cite{vilenkin84}
\begin{equation}
 \theta(z_l,z_s)=\frac{8\pi
  G\mu}{c^2}\frac{D_{ls}}{D_{os}}\sin\phi\equiv
  \delta\frac{D_{ls}}{D_{os}}\sin\phi,
\label{sep}
\end{equation}
where $D_{ls}$ and $D_{os}$ are angular diameter distances between lens
(string) and source (galaxy) and observer and source, respectively, and
$\phi$ is the angle between the string direction and the line-of-sight.
We defined the dimensionless parameter $\delta$ which characterizes a
cosmic string.

Eq. (\ref{sep}) implies that it is difficult to obtain string parameters
such as $\delta$ and the redshift $z_l$, even if we examine a single
string lens system in great details: Assuming the redshift of the lensed 
galaxy is known, it is clear that we can constrain only the combination
of $\delta$, $z_l$, and $\phi$, and there is no way to separate this
constraint. Therefore we have to resort to other methods. In this paper,
we consider the {\it statistics} of string lensing. Below we derive
explicit expressions for the distributions of image separations, and show
that they are quite effective to determine these parameters separately. 

\subsection{Straight String}
First we consider the case that the cosmic string is straight within
the field-of-view we are interested in. As in \cite{huterer03}, we
study the number of other nearby lensing events, given the detection
of one lensing event by a cosmic string. In this case, the number of
lenses with image separations larger than $\theta$, $N(>\theta)$, is
computed as 
\begin{equation}
 N(>\theta)=L\int_{z_l}^\infty
  dz_s\frac{dn_{\rm gal}}{dz_s}\theta_{\rm
  lens}\Theta(\theta(z_l,z_s)-\theta),
\label{n_cum_str}
\end{equation}
where $\Theta(x)$ is the Heviside step function, $dn_{\rm gal}/dz$ is the
number distribution of background galaxies, and $\theta_{\rm lens}$ is
the cross section diameter. The total (projected) length of the
cosmic string is defined by $L$: If the field-of-view is a circle with
radius $R$, it is simply given by $L=2R$. In the specific examples
below, we assume the following form \cite{huterer03} 
\begin{equation}
\frac{dn_{\rm gal}}{dz}=A\frac{4}{\sqrt{\pi}}\left(\frac{z}{z_0}\right)^2
\exp\left[-\left(\frac{z}{z_0}\right)^2\right]\frac{1}{z_0},
\label{ngal}
\end{equation}
where $A$ is the total number density of galaxies. The cross section
diameter $\theta_{\rm lens}$ is simply given by $\theta$ if we consider
only those whose centers are multiply imaged. More strictly, it is
possible that a part of a galaxy is multiply imaged. To compute this, we
adopt an exponential disk $I(r)=\exp(-r/r_{\rm d})$ for the surface
brightness profile of galaxies. We assume $r_{\rm d}=1h^{-1}$kpc
throughout the paper. From this profile, we can compute $\theta_{\rm
lens}$ as a function of limiting flux ratios $f_{\rm lim}$, i.e., 
$\theta_{\rm lens}=\theta_{\rm lens}(f_{\rm lim})$. We note that
$\theta_{\rm lens}(f_{\rm lim}=0.5)=\theta$, if the size of galaxy 
is much smaller than the image separation $\theta$. 

The differential image separation distribution of lensed galaxies is
obtained from Eq. (\ref{n_cum_str})
\begin{equation}
 \frac{dN}{d\theta}=L\int_{z_l}^\infty
  dz_s\frac{dn_{\rm gal}}{dz}\theta_{\rm
  lens}\delta(\theta(z_l,z_s)-\theta),
\label{n_diff_str}
\end{equation}
where $\delta(x)$ is the delta function.

\subsection{Non-straight String}
Next, we consider the case that the coherent length of the string,
$\xi$, is much smaller than the field-of-view we search for lensed
galaxies. Put another way, the cosmic string is not straight but
wiggling within the field-of-view. Again, we study the number of other
nearby lensing events, given the detection of one lensing event by a
cosmic string. In this case, the distributions can be obtained by
averaging Eqs. (\ref{n_cum_str}) and (\ref{n_diff_str}) over the
angle $\phi$ with the weight of $\sin\phi$: 
\begin{equation}
 N(>\theta)=L\int_0^\pi\frac{d\phi}{2}\sin\phi\int_{z_l}^\infty
  dz_s\frac{dn_{\rm gal}}{dz_s}\theta_{\rm
  lens}\Theta(\theta(z_l,z_s)-\theta),
\label{n_cum}
\end{equation}
\begin{equation}
 \frac{dN}{d\theta}=L\int_0^\pi\frac{d\phi}{2}\sin\phi\int_{z_l}^\infty
  dz_s\frac{dn_{\rm gal}}{dz}\theta_{\rm
  lens}\delta(\theta(z_l,z_s)-\theta).
\label{n_diff}
\end{equation}

It is unclear how the string total length $L$ is related with the
coherent length $\xi$. We connect these quantities by assuming a random
walk of the string. In this case, $L$ is given by
\begin{equation}
L=2R\left(\frac{R}{\xi\langle\sin\phi\rangle}\right)=\pi\frac{R^2}{\xi},
\end{equation}
where $R$ is the radius of the field-of-view region. We note that the
change of this relation only affects the overall normalization of the
image separation distribution, and does not change the shape of the
distribution itself.

\section{The Distribution of Image Separations}
\label{sec:dis}

\begin{figure}[!t]
\includegraphics[width=0.45\textwidth]{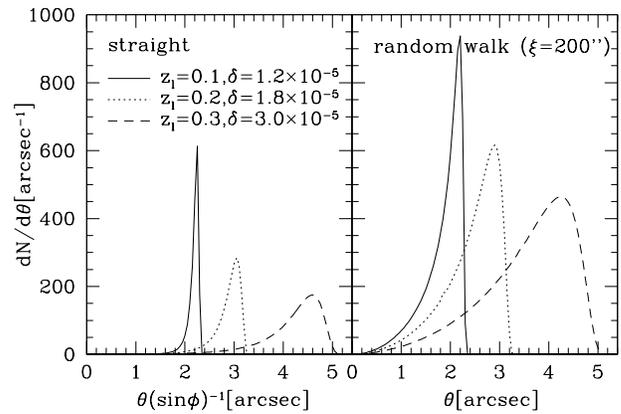}
\caption{The distributions of image separations for the cases of a straight
string ({\it left}) and a wiggling (random walk) string ({\it
  right}). We plot the three cases with different string redshift
$z_l$ and dimensionless tension $\delta$. The total numbers are 
$100$ (solid line), $120$ (dotted line), and $180$ (dashed line) for a
straight string (assuming $\sin\phi=1$), and $240$ (solid line), $310$
(dotted line), and $440$ (dashed line) for a wiggling string. } 
\label{fig:sep}
\end{figure}

We compute the distributions of image separations of lensing events
around one lensing event.  We assume wide-field deep (the limiting
magnitude of $R\sim 28$) imaging by, e.g.,
Suprime-cam on the Subaru telescope \cite{miyazaki02}, and adopt $A=100{\rm
  arcmin^{-1}}$, $R=800''$, and $z_0=1.13$ (from Eq. (\ref{ngal}) this
implies the median galaxy redshift $\langle z\rangle=1.23$). To
illustrate how sensitive the distributions depend on string
parameters, we consider the following three sets of parameters;
($z_l$, $\delta$)=(0.1, $1.2\times10^{-5}$), (0.2, $1.8\times10^{-5}$),
and (0.3, $3.0\times10^{-5}$). Note that these parameter are chosen so
as to be consistent with the possible detection of a cosmic string
\cite{sazhin03} where the image separation is $1.9''$ and the redshift
of lensed galaxy is $z_s=0.46$. We ignore the limiting flux ratio and
adopt $\theta_{\rm lens}=\theta$. We show the result in
Fig. \ref{fig:sep}. We show both straight and random walk cases: For
the latter case, we assumed the correlation length to be $\xi=200''$.
It is clear from this Figure that the shape of the distributions are
quite sensitive to string parameters. This means that the statistics
of lensed galaxies in the field can become a powerful way to constrain
these parameters.

The dependence of the distributions shown in Fig. \ref{fig:sep} on
$\delta$ and $z_l$ can be understood from the following simple
considerations.  First, the maximum image separation increases
monotonically with increasing $\delta$, simply because the maximum
separation is equal to $\delta$ (see Eq. (\ref{sep})). The width of
the distribution is related to the string redshift $z_l$ through a
factor $D_{ls}/D_{os}$ in Eq. (\ref{sep}). The factor rapidly
increases as we increase $z_s$ from $z_l$, and asymptotically
approaches to unity. Therefore, if the string redshift is low
the factor is already close to unity at $z_s\sim 1$ which is typical
redshift of source galaxies, while the factor is still changing at
$z_s\sim 1$ when the string redshift is higher. These explain
qualitative behaviors of the distributions shown in
Fig. \ref{fig:sep}. 

\begin{figure}[!t]
\includegraphics[width=0.45\textwidth]{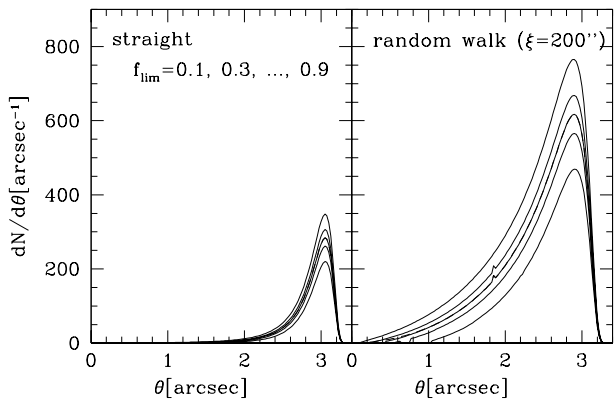}
\caption{The dependence of the distributions on the limiting flux
  ratio $f_{\rm lim}$. For cases of both straight ({\it left}) and
  wiggling ({\it right}) strings, we plot $dN/d\theta$ with five
  different $f_{\rm lim}$: From upper to lower solid lines, we set
  $f_{\rm  lim}=0.1$, $0.3$, $0.5$, $0.7$, and $0.9$.} 
\label{fig:sep_flim}
\end{figure}

Next we study the dependence of the distributions on the limiting flux
ratio $f_{\rm lim}$. The result is shown in Fig. \ref{fig:sep_flim}.
We find that the distributions are not affected very much by $f_{\rm
lim}$. Even if we adopt $f_{\rm lim}=0.9$, the large number of lenses
is still predicted. This implies that a substantial fraction of lensed
pairs should have flux ratios close to unity. 

\section{Constraints on String Parameters}

We showed in the previous section that the distribution of image
separations of galaxies lensed by a cosmic string will be an effective
way to determine the tension and redshift of the cosmic string.
To demonstrate this, we perform likelihood analysis of CSL-1 field
with lens candidates presented by Sazhin et al. \cite{sazhin05}. 
They identified about 9 lens candidates in the $16'\times16'$ field
with CSL-1 at the center. While it is unclear how likely these are
galaxies pairs lensed by a cosmic string because of the lack of
any detailed information on these lens candidates, we adopt 5 lens
candidates with flux ratio larger than $0.5$, as well as CSL-1 itself,
as ``true'' gravitational lens events to see how the model parameters
are constrained by this kind of additional lens events. We assume
wiggling (random walk) cosmic string throughout this section, since
the positions of the lens candidates clearly indicate that the cosmic
string cannot be straight (see \cite{sazhin05}). 

\begin{figure}[!t]
\includegraphics[width=0.45\textwidth]{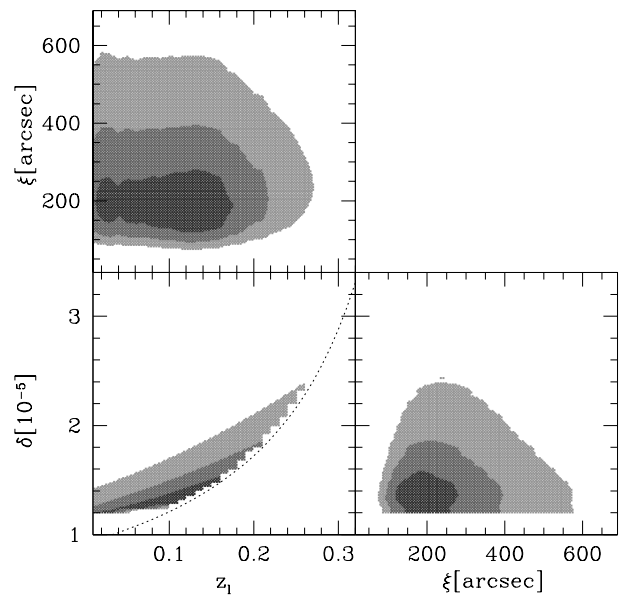}
\caption{Constraints projected in the 2 parameter subspace from the
  3 parameter space, assuming 6 lens candidates in \cite{sazhin05} are
  true lens events. From darker to lighter shaded regions, we show
  $68\%$, $95\%$, $99.7\%$ confidence limits. The dotted line in the
  $z_l$-$\delta$ plane indicates the limit of the prior given by
  Eq. (\ref{pri}).  } 
\label{fig:chi}
\end{figure}

We compute the number distribution of lenses with Eqs. (\ref{n_cum}) and
(\ref{n_diff}). The lens candidates are identified in the medium deep
images with limiting magnitude of $R\sim 24$, which correspond to
$A=10{\rm arcmin^{-1}}$ and $z_0=0.78$ (the median galaxy redshift
$\langle z\rangle=0.85$). From this, we compute the likelihood as
\begin{equation}
{\mathcal L}=\frac{N^x\exp(-N)}{x!}\prod_{i=1}^x\left.\frac{dN}{d\theta}\right|_{\theta_i},
\end{equation}
where $N=N(>1'')$ is the total number of lenses with image separation
larger than $1''$ (lenses with smaller image separations are quite
hard to be identified) expected in the field, $x=6$ is the observed
number of lenses, $\theta_i$ is the image separation of $i$-th
lens. We compute likelihoods as a function of three parameters; the
string redshift $z_l$, the dimensionless tension $\delta$, and the
correlation length $\xi$, and derive constraints on these
parameters. In doing so, we include the following prior 
\begin{equation}
1.9''=9.2\times
  10^{-6}<\delta\left.\frac{D_{ls}}{D_{os}}\right|_{z_s=0.46},
\label{pri}
\end{equation}
which originates from the fact that the redshift of CSL-1 is $z_s=0.46$. 

The resulting constraints are shown in Fig. \ref{fig:chi}. 
The best-fit parameters are $z_l=0.13$, $\delta=1.4\times 10^{-5}$, and 
$\xi=180''$ (which corresponds to $0.3h^{-1}{\rm Mpc}$ at $z=0.13$).
We find that we can place tight constraints on string parameters even
from 6 lens events only: $z_l<0.22$ and $1.2<\delta/10^{-5}<1.8$ at $95\%$
confidence limit. Therefore we conclude that the statistics of lensing
by a cosmic string are useful way to constrain properties of the
cosmic string. Assuming the best-fit parameter set, we predict that
the wide-field imaging of the field (see Sec. \ref{sec:dis} for the
specific setup) will be able to detect $\sim 270$ lens candidates with
the image separations larger than $1''$. 

\section{Conclusions}

In this paper, we derived explicit expressions for the distributions
of image separations of lensing by a cosmic string. We computed the number
of nearby lensing events around one lensing event by a cosmic string.
We considered both cases that the string is straight and fully wiggling
in the field-of-view we are interested in. 

It is quite hard to characterize a cosmic string from detailed
investigation of a single lensing event, because of the degeneracy
between parameters. We have found that the statistics of lensing
events, i.e., comparing the theoretical number distribution of image
separations  with observations is an effective way to break the
degeneracy and determine string parameters such as the string redshift
and tension. We have shown that the detection of only several nearby
lensing events is enough to place interesting new constraints on
parameters. Our calculation indicates that a deep wide-field imaging
in the field of string lens event will be able to find hundreds of
additional lensing events, which offer valuable information on the
cosmic string and hence the very early phase of our universe.

In this paper, we have assumed that the cosmic string is
non-relativistic, i.e., we neglected the separation-angle correction
from its velocity. Statistical study including the velocity effect
would be challenging and necessary for the precise interpretation of
the future observational data. 

\begin{acknowledgments}
We thank J. Polchinski, P. J. Steinhardt and A. J. Tolley for
helpful discussions. K. T. is supported by Grant-in-Aid for JSPS Fellows.
\end{acknowledgments}


\begin{thebibliography}{0}

\bibitem{pogosian03}
L. Pogosian, S. H. H. Tye, I. Wasserman and M. Wyman,
Phys. Rev. D, {\bf 68}, 023506 (2003).

\bibitem{pogosian04}
L. Pogosian, M. Wyman and I. Wasserman, 
JCAP, {\bf 09}, 008 (2004).

\bibitem{jeong05}
E. Jeong and G. F. Smoot, 
Astrophys. J., {\bf 624}, 21 (2005).

\bibitem{lo05}
A. S. Lo and E. L. Wright, 
astro-ph/0503120.

\bibitem{kibble04}
T. W. B. Kibble, 
astro-ph/0410073.

\bibitem{polchinski04}
J. Polchinski, 
hep-th/0412244.

\bibitem{perivolaropoulos05}
L. Perivolaropoulos, 
astro-ph/0501590.

\bibitem{jones03}
N. T. Jones, H. Stoica and S.-H. H. Tye, 
Phys. Lett. B, {\bf 563}, 6 (2003).

\bibitem{dvali04}
G. Dvali and A. Vilenkin, 
JCAP, {\bf 0403}, 010 (2004).

\bibitem{copeland04}
E. J. Copeland, R. C. Myers and J. Polchinski, 
JHEP, {\bf 0406}, 013 (2004).

\bibitem{jackson04}
M. G. Jackson, N. T. Jones and J. Polchinski, 
hep-th/0405229.

\bibitem{jeannerot03}
R. Jeannerot, J. Rocher and M. Sakellariadou, 
Phys. Rev. D {\bf 68}, 103514 (2003).

\bibitem{sazhin03}
M. Sazhin {\it et al.},
Mon. Not. Roy. Astron. Soc., {\bf 343}, 353 (2003); 
M. Sazhin {\it et al.},
astro-ph/0506400.

\bibitem{donaire05}
M. Donaire and A. Rajantie,
hep-ph/0508272.

\bibitem{huterer03}
D. Huterer and T. Vachaspati, 
Phys. Rev. D, {\bf 68}, 041301 (2003).

\bibitem{shlaer05}
B. Shlaer and S. -H. H. Tye, 
Phys. Rev. D, {\bf 72}, 043532 (2005).

\bibitem{vilenkin84}
A. Vilenkin, 
Astrophys. J., {\bf 282}, L51 (1984).

\bibitem{miyazaki02}
S. Miyazaki {\it et al.}, 
Publ. Astron. Soc. Jap., {\bf 54}, 833 (2002).

\bibitem{sazhin05}
M. Sazhin {\it et al.},
Astron. Lett., {\bf 31}, 73 (2005).

\end{thebibliography}
\end{document}